\documentclass[conference]{IEEEtran}
\IEEEoverridecommandlockouts
\usepackage{indentfirst}
\usepackage[dvips]{graphicx}
\usepackage{amsfonts}
\usepackage{multirow}
\usepackage{amsmath,amsthm,amssymb}
\usepackage{color,changepage}
\usepackage{algorithm}
\usepackage{algorithmic}
\usepackage{bbm}
\usepackage{verbatim}
\usepackage{makecell}
\usepackage{subfigure}
\usepackage{mathrsfs}
\usepackage{arydshln}
\usepackage{cases}
\usepackage{threeparttable}
\usepackage{extarrows}
\usepackage{array}
\usepackage{bm}
\usepackage{pifont}
\usepackage[T1]{fontenc}
\usepackage{threeparttable}
\usepackage{booktabs}
\usepackage{mdwlist}
\usepackage{hyperref}
\usepackage{cite}
\allowdisplaybreaks[4]

\definecolor{newcolor}{rgb}{0.5,0,1}

\newcolumntype{I}{!{\vrule width 2pt}}

\newtheorem{Theorem}{Theorem}

\newtheorem{Lemma}{Lemma}

\theoremstyle{remark}
\newtheorem{Remark}{Remark}

\DeclareMathAlphabet{\mathpzc}{OT1}{pzc}{m}{it}

\definecolor{newcolor}{rgb}{0.5,0,1}

\def\BibTeX{{\rm B\kern-.05em{\sc i\kern-.025em b}\kern-.08em
    T\kern-.1667em\lower.7ex\hbox{E}\kern-.125emX}}

\begin{document}

\title{
Private Multiple Linear Computation: A Flexible  Communication-Computation Tradeoff
}
\author{
\IEEEauthorblockN{Jinbao Zhu, Lanping Li, Xiaohu Tang, and Ping Deng}
\IEEEauthorblockA{
Southwest Jiaotong University, Chengdu 611756, China\\
E-mail: jinbaozhu@swjtu.edu.cn, lilanping523@gmail.com, xhutang@swjtu.edu.cn, pdeng@swjtu.edu.cn
}}
\maketitle

\begin{abstract}
We consider the problem of private multiple linear computation (PMLC) over a replicated storage system with colluding and unresponsive constraints. In this scenario, the user wishes to privately compute $P$ linear combinations of $M$ files from a set of $N$ replicated servers without revealing any information about the coefficients of these linear combinations to any $T$ colluding servers, in the presence of $S$ unresponsive servers that do not provide any information in response to user queries.
Our focus is on more general performance metrics where the communication and computational overheads incurred by the user are not neglected. Additionally, the communication and computational overheads for servers are also taken into consideration.
Unlike most previous literature that primarily focused on download cost from servers as a performance metric, we propose a novel PMLC scheme to establish a flexible tradeoff between communication costs and computational complexities.
%
\end{abstract}

\section{Introduction}\label{Introduction}

\IEEEPARstart{W}{ith} the rapid evolution of big data, machine learning algorithms, and artificial intelligence models, distributed systems have emerged as a promising solution to compute large-scale data. Nevertheless, scaling out computation to distributed systems inevitably brings up additional challenges in preserving computational privacy. Recent studies have extensively addressed the application of coding-theoretic tools to safeguard computational privacy in distributed environments, including areas such as private search \cite{ostrovsky2005private,chen2020asymptotic}, private linear transformation \cite{heidarzadeh2022single,kazemi2021multi}, and secure matrix multiplication \cite{chang2018capacity,d2020gasp,aliasgari2020private,zhu2021improved,li2022private,zhu2022systematic}. Among these, the problem of private linear computation (PLC) has garnered widespread attention in the past few years.

The PLC problem was initially introduced in \cite{sun2018capacity,mirmohseni2018private}. In the classical PLC setting, a user wishes to compute a linear combination of $M$ files from a distributed system with $N$ servers, without revealing anything about the coefficients of the linear combination to any individual server, where the $M$ files, each of length $L$, are stored at the $N$ servers in a repetition form. To this end, the user sends a private query to each server, which accordingly responds to an answer. Ultimately, the user can decode the desired linear computation from the collected answers. Indeed, the PLC problem is a linear generalization of private information retrieval (PIR) wherein the user aims to privately retrieve one out of the files from the servers, while just hiding the identity of the desired file. For further details on PIR, refer to \cite{chor1998private,gasarch2004survey,sun2017capacity,tian2019capacity,banawan2018capacity,tajeddine2018private,zhu2019new,zhou2020capacity,sun2017capacitycolluding,zhang2018optimal,yao2021capacity,freij2017private,tajeddine2019private,banawan2018multi,jia2019cross,zhu2022multi,jia2022x,li2020towards,ulukus2022private,vithana2023private}.

For the PLC problem, the optimal download cost from the servers was exactly characterized as $\frac{(1-(1/N)^M)L}{1-1/N}$ in \cite{sun2018capacity}.
Subsequently, the problem of PLC over maximum distance separable (MDS) coded storage system, where the files are distributed across the $N$ servers according to $(N, K')$ MDS codes, was considered in \cite{obead2022private,obead2018capacity,obead2018achievable} and its optimal download cost was proven to be $\frac{(1-({K'}/{N})^M)L}{1-{K'}/{N}}$.
Reference \cite{jia2023asymptotic} focused on a more general PLC problem with graph-based replicated storage and characterized its asymptotic optimal download cost in the limit as the number of files $M$ approaches infinity, where every file is secretly stored among a specific subset of servers, and the coefficients of desired linear combination is private against a certain number of colluding servers. 
In a recent work \cite{gholami2023multi}, the problem of non-colluding private multiple linear computation (PMLC) with replicated storage was considered to privately compute multiple linear combinations, and its goal is only to minimize the download cost. The nonlinear extension of PLC, private polynomial computation, was explored in \cite{raviv2019private,zhu2022symmetric,obead2022privatePPC}, where the user aims to privately compute a multivariate polynomial function of the files.

\vspace{-0.1mm}

This paper considers the $T$-colluding PMLC problem over a replicated storage system with unresponsive constraint, i.e., the user wishes to compute $P$ linear combinations of the $M$ files from $N$ replicated servers, while guaranteeing that any $T$ colluding servers learn nothing about the coefficients of these linear combinations, in the presence of $S$ unresponsive servers that do not respond any information at all.
To the best of our knowledge, all current PLC \cite{sun2018capacity,mirmohseni2018private,obead2022private,obead2018capacity,obead2018achievable,jia2023asymptotic} and PMLC \cite{gholami2023multi} works focus solely on minimizing the download cost as a performance metric. 
As is well-known, in practical communication networks, uplink communication is more expensive and time-consuming than downlink communication. Therefore, it is important to consider the upload cost of queries as a performance metric.
Particularly, due to privacy constraint, the upload cost of queries and the computational complexity of generating queries are scaled with the number of files $M$, and thus they are often significant and should not be neglected, especially when $M$ is large.
Moreover, the computational complexity of generating answers and the decoding complexity for recovering desired computations should also be taken into consideration, since they are scaled with the file length $L$.

\vspace{-0.1mm}

The main contribution of this paper is to propose a novel PMLC scheme that establishes a flexible tradeoff between communication costs (upload and download) and computational complexities (queries, answers, and decoding) for the PMLC problem, which is more efficient in optimizing system performance to accelerate private computation.



\subsubsection*{Notation}
Throughout this paper, all indices start from $0$, including matrix indices. 
Let boldface capital and lower-case letters represent matrices and vectors, respectively, e.g., $\mathbf{C}$ and $\mathbf{w}$.
Let cursive capital letters denote sets, such as $\mathcal{S}$, while $|\mathcal{S}|$ denotes its cardinality. 
For any integers $m,n$ such that $m<n$, $[m:n)$ denotes the set $\{m,m+1,\ldots,n-1\}$. 
Define ${A}_{\mathcal{S}}$ as $\{{A}_{s_0},{A}_{s_1},\!\ldots\!,{A}_{s_{n\!-\!1}}\}$ for any index set $\mathcal{S}=\{s_0,s_1,\ldots,s_{n\!-\!1}\}$ of size $n$.
Denote by $(\cdot)_n$ the modulo $n$ operation.

\section{Problem Formulation}
Consider $M$ independent files $\mathbf{w}^{(0)},\ldots,\mathbf{w}^{(M-1)}$, each of which consists of $L$ symbols drawn independently and uniformly from a finite field $\mathbb{F}_q$ for a prime power $q$. 

There are $N$ distributed servers and each server stores all the $M$ files $\mathbf{w}^{(0)},\ldots,\mathbf{w}^{(M-1)}$. 
A user wishes to compute $P$ linear combinations of the $M$ files from the $N$ servers while keeping the coefficients of these $P$ linear combinations private from any group of up to $T$ colluding servers. 
We assume the presence of some servers $\mathcal{S}$ of size at most $S$ that do not respond at all when the user accesses these servers, known as \emph{unresponsive servers}. Notably, their identities are unknown in advance and may change with time.

Denote the coefficients of the $P$ linear combinations by a matrix $\mathbf{C}$ of dimensions $P\times M$ with all elements being distributed on $\mathbb{F}_q$, given by 
\begin{IEEEeqnarray}{rCl}\label{coefficient:matrix}
\mathbf{C} =\left[
\begin{array}{@{\;}ccc@{\;}}
c_{0,0} &  \ldots & c_{0,M-1}  \\
 \vdots &  \ddots & \vdots \\
c_{P-1,0} &  \ldots &c_{P-1,M-1}\\
\end{array}
\right].
\end{IEEEeqnarray}
Formally, the user will employ a private multiple linear computation (PMLC) scheme consisting of the following  phases:


\begin{enumerate}
\item \emph{Query Phase:} The user locally generates $N$ queries $\mathcal{Q}^{(\mathbf{C})}_{[0:N)}$ and sends $\mathcal{Q}^{(\mathbf{C})}_{n}$ to server $n$ for all $n\in[0:N)$. 
\item \emph{Answer Phase:} 
Upon receiving the query $\mathcal{Q}_{n}^{(\mathbf{C})}$, each unresponsive server $n$ in $\mathcal{S}$ will not provide any information, and each remaining server $n$ in $[0:N)\backslash\mathcal{S}$ will respond an answer $\textbf{A}_{n}^{(\textbf{C})}$ according to the received query and the stored $M$ files $\mathbf{w}_{0},\ldots,\mathbf{w}_{M-1}$, for any $\mathcal{S}\subseteq[0:N)$ of size at most $S$.
\item \emph{Decoding Phase:} The user waits for the responses from any $N-S$ servers, and then recovers the $P$ desired linear computations 
\begin{IEEEeqnarray}{c}\label{desired:computation}
\widetilde{\textbf{w}}^{(i)}\!=\! c_{i,0}\textbf{w}^{(0)}\!+\!\ldots\!+\!c_{i,M-1}\textbf{w}^{(M-1)},\;\;\;\forall\,i\!\in\![0\!:\!P). \IEEEeqnarraynumspace
\end{IEEEeqnarray}
\end{enumerate}
A valid PMLC scheme must satisfy the following constraints.
\begin{itemize}
\item \textbf{Correctness:} The user must be able to correctly decode $\widetilde{\textbf{w}}^{(0)},\ldots,\widetilde{\textbf{w}}^{(P-1)}$ from the local queries and the received responses from any $N-S$ servers, i.e., for any $\mathcal{S}\subseteq[0:N)$ with $|\mathcal{S}|\leq S$, 
\begin{IEEEeqnarray}{c}\notag
H(\widetilde{\textbf{w}}^{(0)},\ldots,\widetilde{\textbf{w}}^{(P-1)}|\mathbf{A}_{[0:N)\backslash\mathcal{S}}^{(\mathbf{C})},\mathcal{Q}_{[0:N)}^{(\mathbf{C})})=0.
\end{IEEEeqnarray}
\item \textbf{Privacy:} The queries sent to any up to $T$ colluding servers reveal nothing about the coefficients $\mathbf{C}$ of the $P$ linear combinations, i.e., for any $\mathcal{T}\!\subseteq\![0\!:\!N)$ with $|\mathcal{T}|\!\leq\!T$, 
\begin{IEEEeqnarray}{c}\label{Infor:priva cons}
I(\mathbf{C};\mathcal{Q}_{\mathcal{T}}^{(\mathbf{C})})=0.
\end{IEEEeqnarray}
\end{itemize}

In general, the performance of a PMLC scheme is evaluated by the following two metrics.
\begin{enumerate}
\item[1.] The communication costs, which are comprised of the upload cost for the queries sent to the $N$ servers and the download cost from any $N-S$ servers, defined as
\begin{IEEEeqnarray}{c}
U\!=\!\sum\limits_{n=0}^{N-1}\!H(\mathcal{Q}_n^{(\mathbf{C})}), \;\; D\!=\!\max\limits_{\substack{\mathcal{S}\subseteq[0:N)\\|\mathcal{S}|=S}}\sum\limits_{n\in[0:N)\backslash\mathcal{S}}\!\!H(\mathbf{A}_{n}^{(\mathbf{C})}).\notag
\end{IEEEeqnarray}
\item[2.]  The computational complexities, which include the complexities of queries, server computation, and decoding. The query complexity at the user is defined as the order of the number of arithmetic operations required to generate all the queries $\mathcal{Q}_{[0:N)}^{(\mathbf{C})}$. The server computation complexity is defined as the order of the number of arithmetic operations required to generate the response 
$\mathbf{A}_n^{(\mathbf{C})}$, maximized over $n\in[0:N)$. The decoding complexity at the user is defined as the number of arithmetic operations required to decode the desired linear computations $\widetilde{\textbf{w}}^{(0)},\ldots,\widetilde{\textbf{w}}^{(P-1)}$ from the answers of responsive servers.
\end{enumerate}

The objective of this paper is to design efficient PMLC schemes that minimize communication costs and computational complexities.

\section{Main Result and Discussions}\label{Main:Result}
\begin{Theorem}\label{PMLC:theorem}
For the $T$-colluding PMLC problem with $M$ files of each length $L$, $P$ desired linear combinations, and $N$ replicated servers among which $S$ are unresponsive, the PMLC scheme proposed in Section \ref{PMLC:scheme} achieves the following metrics for any given integer parameters $K>0,E>0,R\geq 0$ such that $K+R\leq N-S-T,E|L,N|ME$, and $K|PE$.\footnote{As in current big data era, the file length $L$ and the number $M$ of files are typically large. For convenience, we assume $E|L$ and $N|ME$.} 
\begin{IEEEeqnarray*}{l}
\text{Upload Cost:}\;\;   \frac{(N-R)E^2MP}{K},\\
\text{Download Cost:}\;\;  \frac{(N-S)PL}{K}, \\
\text{Query Complexity:}\;\;\\ \quad\;\; O\!\left(\!\frac{E^2MP(N\!-\!R)(\log(N\!-\!R))^2\log\log(N\!-\!R)}{K}\!\right), \\
\text{Server Computation Complexity:}\;\; O\!\left(\!\frac{E(N\!-\!R)MPL}{NK}\!\right), \\
\text{Decoding Complexity:}\\
\quad\quad\quad\; O\!\left(\!\frac{PL(N\!-\!S)(\log (N\!-\!S))^2\log\log (N\!-\!S)}{K}\!\right).
\end{IEEEeqnarray*}
\end{Theorem}
The proposed PMLC scheme achieves a flexible tradeoff between communication costs and computational complexities by adjusting the hyperparameters $K,E,R$ satisfying $K+R\leq N-S-T$ and $K|PE$. More specifically, from the theorem, it is straightforward to observe that
\begin{itemize}
\item As the parameter $E$ increases, the download cost and decoding complexity remain unchanged, while the upload cost, query complexity, and server computation complexity increase. Thus, $E$ should be minimized under the constraint $K|PE$ and accordingly $E$ is set to ${K}/{\gcd(K,P)}$.
\item The download cost and decoding complexity are independent of the parameter $R$, while the upload cost, query complexity, and server computation complexity decrease as $R$ increases.
\item All the performance metrics decrease as $K$ increases.
It is highly valuable to point out that due to the constraints $K+R\leq N-S-T$ and $K|PE$, increasing $K$ may not necessarily be the best choice. For example, consider the traditional private linear computation (PLC) scenario (i.e., $P=1$) \cite{sun2018capacity,mirmohseni2018private} with the minimum value of $E$ being $K$. 
In this case, the server computation complexity is independent of $K$ and decreases with increasing $R$, while the upload cost and query complexity increase with increasing $K$ and decrease with increasing $R$. Additionally, the download cost and decoding complexity decrease with increasing $K$ and are independent of $R$.
Due to $K+R\leq N-S-T$, $R$ should be maximized if server computation complexity, upload cost, and query complexity are the primary metrics; conversely, $K$ should be maximized if download cost and decoding complexity are the primary performance metrics. 
\end{itemize}
In practice, one can choose appropriate $K,E,R$ according to the actual system parameters $P,M,L$ and the available system resources, such as the communication and computation capabilities of the user and servers.

\begin{Remark}
The PMLC problem considered in this paper includes the traditional $T$-colluding PLC  problem with $S$ unresponsive servers as a special case by setting $P=1$.
For this PLC problem, a lower bound of the optimal download cost is proven to be $\frac{(1-({T}/{(N-S)})^M)L}{1-{T}/{(N-S)}}$ \cite{sun2017capacitycolluding}, which converges exponentially to its asymptotic value $\frac{(N-S)L}{N-S-T}$ with the number of files $M$. In the special case of $P=1,R=0$, and $K=E=N-S-T$, our PMLC scheme achieves the download cost $\frac{(N-S)L}{N-S-T}$. That is, our PMLC scheme can degrade to a PLC scheme with the asymptotic optimal download cost as the number of files approaches infinity.
\end{Remark}


\section{A Flexible Private Multiple Linear Computation Scheme}\label{PMLC:scheme}
In this section, we formally describe the proposed PMLC scheme and analyze its communication costs, computational complexities, and privacy guarantee. This completes the proof of Theorem \ref{PMLC:theorem}.


We begin with a simple example to illustrate the key ideas underlying the proposed PMLC scheme. 

\subsection{Illustrative Example}
Consider the PMLC problem with system parameters $N=6,T=1,S=1,M=3,P=3,L=4$, i.e., the user wishes to compute $3$ linear combinations of the $3$ files $\mathbf{w}^{(0)},\mathbf{w}^{(1)},\mathbf{w}^{(2)}$, each of length $4$, from $6$ replicated servers with $1$-privacy and $1$-unresponsiveness constraints. 

Denote the $3$ files $\mathbf{w}^{(0)},\mathbf{w}^{(1)},\mathbf{w}^{(2)}$ by
\begin{IEEEeqnarray}{rCl}
\mathbf{w}^{(0)}&=&\big(w^{(0)}_0,w^{(0)}_1,w^{(0)}_2,w^{(0)}_3\big), \label{example:file1} \\
\mathbf{w}^{(1)}&=&\big(w^{(1)}_0,w^{(1)}_1,w^{(1)}_2,w^{(1)}_3\big), \\
\mathbf{w}^{(2)}&=&\big(w^{(2)}_0,w^{(2)}_1,w^{(2)}_2,w^{(2)}_3\big).
\end{IEEEeqnarray}
The coefficients of the $3$ linear combinations are given by
\begin{IEEEeqnarray}{c}\label{example:coefficients}
\mathbf{C} =\left[
\begin{array}{@{\;}ccc@{\;}}
c_{0,0} & c_{0,1} & c_{0,2}  \\
c_{1,0} & c_{1,1} & c_{1,2}  \\
c_{2,0} & c_{2,1} & c_{2,2}  \\
\end{array}
\right].
\end{IEEEeqnarray}

Let us choose the parameters $K=3, R=1, E=2$.\footnote{Here, we choose these values solely to better convey the ideas behind the proposed PMLC scheme.} Then these three files stored at each server can be rearranged into a matrix $\mathbf{W}$ of dimensions $ME\times{L}/{E}=6\times 2$ as follow.
\begin{IEEEeqnarray}{c}\label{example:filepartition}
\mathbf{W} =\left[
\begin{array}{@{\;}c@{\;}}
\mathbf{w}_0 \\
\mathbf{w}_1 \\
\mathbf{w}_2 \\
\mathbf{w}_3 \\
\mathbf{w}_4 \\
\mathbf{w}_5 \\
\end{array}
\right]
=\left[
\begin{array}{@{\;}cc@{\;}}
w^{(0)}_0 & w^{(0)}_1 \\
w^{(1)}_0 & w^{(1)}_1 \\
w^{(2)}_0 & w^{(2)}_1 \\
w^{(0)}_2 & w^{(0)}_3 \\
w^{(1)}_2 & w^{(1)}_3 \\
w^{(2)}_2 & w^{(2)}_3 \\
\end{array}
\right],
\end{IEEEeqnarray}
where $\mathbf{w}_i, i=0,1,2,3,4,5$ is the $i$-th row of the matrix $\mathbf{W}$.
Furthermore, we expand the coefficient matrix $\mathbf{C}$ into a matrix $\widetilde{\mathbf{C}}$ of dimensions $PE\times ME=6\times 6$, given by
\begin{IEEEeqnarray}{c}\label{example:partition}
\widetilde{\mathbf{C}} =\left[
\begin{array}{@{\;}cccccc@{\;}}
c_{0,0} & c_{0,1} & c_{0,2} & 0 & 0 & 0  \\
c_{1,0} & c_{1,1} & c_{1,2} & 0 & 0 & 0  \\ \hdashline[1.5pt/1pt]
c_{2,0} & c_{2,1} & c_{2,2} & 0 & 0 & 0  \\
0 & 0 & 0 & c_{0,0} & c_{0,1} & c_{0,2}  \\ \hdashline[1.5pt/1pt]
0 & 0 & 0 & c_{1,0} & c_{1,1} & c_{1,2}  \\
0 & 0 & 0 & c_{2,0} & c_{2,1} & c_{2,2}  \\
\end{array}
\right].
\end{IEEEeqnarray}
Afterwards, we horizontally partition $\widetilde{\mathbf{C}}$ into $K=3$ blocks, such that each column of $\widetilde{\mathbf{C}}$ comprises $3$ column vectors, each of length $PE/K=2$, as indicated by the dashed lines in \eqref{example:partition}. 
The user can easily accomplish the $3$ desired linear computations by obtaining the results $\widetilde{\mathbf{C}}\cdot\mathbf{W}$.

Let $\{\beta_{0},\beta_{1},\beta_{2},\beta_3,\alpha_{0},\alpha_1,\alpha_2,\alpha_3,\alpha_4,\alpha_5\}$ be $N+K+T=10$ distinct elements from $\mathbb{F}_q$.
To provide $1$-privacy guarantee, we employ $1$ random noise vector to mask the $3$ partitioning coefficient vectors in each column of $\widetilde{\mathbf{C}}$. This is achieved using a Lagrange interpolating polynomial. Particularly, to establish a more flexible performance tradeoff, we additionally force the evaluation of this polynomial to be a zero vector at $R=1$ specific point.
Specifically, let $z_{0}^{0},\ldots,z_1^5$ be random noises, and then construct an interpolating polynomial of degree at most $K+T+R-1=4$ for every column of $\widetilde{\mathbf{C}}$, satisfying
\begin{IEEEeqnarray}{rClrClrClrClrCl}
f_{\!0}\!(\!\beta_0\!)\!\!&=&\!\!\!\left[
\begin{array}{@{\!}c@{\!}}
c_{0,0} \\
c_{1,0} \\
\end{array}
\right]\!\!,\;&
f_{\!0}\!(\!\beta_1\!)\!\!&=&\!\!\!\left[
\begin{array}{@{\!}c@{\!}}
c_{2,0} \\
0 \\
\end{array}
\right]\!\!,\;&
f_{\!0}\!(\!\beta_2\!)\!\!&=&\!\!\!\left[
\begin{array}{@{}c@{}}
0 \\
0 \\
\end{array}
\right]\!\!,\;&
f_{\!0}\!(\!\beta_3\!)\!\!&=&\!\!\!\left[
\begin{array}{@{}c@{}}
z_{0}^{0} \\
z_{1}^{0} \\
\end{array}
\right]\!\!,\;&
f_{\!0}\!(\!\alpha_0\!)\!\!&=&\!\!\!\left[
\begin{array}{@{}c@{}}
0 \\
0 \\
\end{array}
\right]\!\!,\notag \\
f_{\!1}\!(\!\beta_0\!)\!\!&=&\!\!\!\left[
\begin{array}{@{\!}c@{\!}}
c_{0,1} \\
c_{1,1} \\
\end{array}
\right]\!\!,\;&
f_{\!1}\!(\!\beta_1\!)\!\!&=&\!\!\!\left[
\begin{array}{@{\!}c@{\!}}
c_{2,1} \\
0 \\
\end{array}
\right]\!\!,\;&
f_{\!1}\!(\!\beta_2\!)\!\!&=&\!\!\!\left[
\begin{array}{@{}c@{}}
0 \\
0 \\
\end{array}
\right]\!\!,\;&
f_{\!1}\!(\!\beta_3\!)\!\!&=&\!\!\!\left[
\begin{array}{@{}c@{}}
z_{0}^{1} \\
z_{1}^{1} \\
\end{array}
\right]\!\!,\;&
f_{\!1}\!(\!\alpha_1\!)\!\!&=&\!\!\!\left[
\begin{array}{@{}c@{}}
0 \\
0 \\
\end{array}
\right]\!\!,\notag\\
f_{\!2}\!(\!\beta_0\!)\!\!&=&\!\!\!\left[
\begin{array}{@{\!}c@{\!}}
c_{0,2} \\
c_{1,2} \\
\end{array}
\right]\!\!,\;&
f_{\!2}\!(\!\beta_1\!)\!\!&=&\!\!\!\left[
\begin{array}{@{\!}c@{\!}}
c_{2,2} \\
0 \\
\end{array}
\right]\!\!,\;&
f_{\!2}\!(\!\beta_2\!)\!\!&=&\!\!\!\left[
\begin{array}{@{}c@{}}
0 \\
0 \\
\end{array}
\right]\!\!,\;&
f_{\!2}\!(\!\beta_3\!)\!\!&=&\!\!\!\left[
\begin{array}{@{}c@{}}
z_{0}^{2} \\
z_{1}^{2} \\
\end{array}
\right]\!\!,\;&
f_{\!2}\!(\!\alpha_2\!)\!\!&=&\!\!\!\left[
\begin{array}{@{}c@{}}
0 \\
0 \\
\end{array}
\right]\!\!,\notag\\
f_{\!3}\!(\!\beta_0\!)\!\!&=&\!\!\!\left[
\begin{array}{@{}c@{}}
0 \\
0 \\
\end{array}
\right]\!\!,\;&
f_{\!3}\!(\!\beta_1\!)\!\!&=&\!\!\!\left[
\begin{array}{@{\!}c@{\!}}
0\\
c_{0,0} \\
\end{array}
\right]\!\!,\;&
f_{\!3}\!(\!\beta_2\!)\!\!&=&\!\!\!\left[
\begin{array}{@{\!}c@{\!}}
c_{1,0} \\
c_{2,0} \\
\end{array}
\right]\!\!,\;&
f_{\!3}\!(\!\beta_3\!)\!\!&=&\!\!\!\left[
\begin{array}{@{}c@{}}
z_{0}^{3} \\
z_{1}^{3} \\
\end{array}
\right]\!\!,\;&
f_{\!3}\!(\!\alpha_3\!)\!\!&=&\!\!\!\left[
\begin{array}{@{}c@{}}
0 \\
0 \\
\end{array}
\right]\!\!,\notag\\
f_{\!4}\!(\!\beta_0\!)\!\!&=&\!\!\!\left[
\begin{array}{@{}c@{}}
0 \\
0 \\
\end{array}
\right]\!\!,\;&
f_{\!4}\!(\!\beta_1\!)\!\!&=&\!\!\!\left[
\begin{array}{@{\!}c@{\!}}
0\\
c_{0,1} \\
\end{array}
\right]\!\!,\;&
f_{\!4}\!(\!\beta_2\!)\!\!&=&\!\!\!\left[
\begin{array}{@{\!}c@{\!}}
c_{1,1} \\
c_{2,1} \\
\end{array}
\right]\!\!,\;&
f_{\!4}\!(\!\beta_3\!)\!\!&=&\!\!\!\left[
\begin{array}{@{}c@{}}
z_{0}^{4} \\
z_{1}^{4} \\
\end{array}
\right]\!\!,\;&
f_{\!4}\!(\!\alpha_4\!)\!\!&=&\!\!\!\left[
\begin{array}{@{}c@{}}
0 \\
0 \\
\end{array}
\right]\!\!,\notag\\
f_{\!5}\!(\!\beta_0\!)\!\!&=&\!\!\!\left[
\begin{array}{@{}c@{}}
0 \\
0 \\
\end{array}
\right]\!\!,\;&
f_{\!5}\!(\!\beta_1\!)\!\!&=&\!\!\!\left[
\begin{array}{@{\!}c@{\!}}
0\\
c_{0,2} \\
\end{array}
\right]\!\!,\;&
f_{\!5}\!(\!\beta_2\!)\!\!&=&\!\!\!\left[
\begin{array}{@{\!}c@{\!}}
c_{1,2} \\
c_{2,2} \\
\end{array}
\right]\!\!,\;&
f_{\!5}\!(\!\beta_3\!)\!\!&=&\!\!\!\left[
\begin{array}{@{}c@{}}
z_{0}^{5} \\
z_{1}^{5} \\
\end{array}
\right]\!\!,\;&
f_{\!5}\!(\!\alpha_5\!)\!\!&=&\!\!\!\left[
\begin{array}{@{}c@{}}
0 \\
0 \\
\end{array}
\right]\!\!.\notag
\end{IEEEeqnarray}
Then the queries are designed by evaluating these polynomials:
\begin{IEEEeqnarray*}{rCl}
\mathcal{Q}_0^{(\mathbf{C})}&=&\big\{f_1(\alpha_0),f_2(\alpha_0),f_3(\alpha_0),f_4(\alpha_0),f_5(\alpha_0) \big\},\\
\mathcal{Q}_1^{(\mathbf{C})}&=&\big\{f_0(\alpha_1),f_2(\alpha_1),f_3(\alpha_1),f_4(\alpha_1),f_5(\alpha_1) \big\},\\
\mathcal{Q}_2^{(\mathbf{C})}&=&\big\{f_0(\alpha_2),f_1(\alpha_2),f_3(\alpha_2),f_4(\alpha_2),f_5(\alpha_2) \big\},\\
\mathcal{Q}_3^{(\mathbf{C})}&=&\big\{f_0(\alpha_3),f_1(\alpha_3),f_2(\alpha_3),f_4(\alpha_3),f_5(\alpha_3) \big\},\\
\mathcal{Q}_4^{(\mathbf{C})}&=&\big\{f_0(\alpha_4),f_1(\alpha_4),f_2(\alpha_4),f_3(\alpha_4),f_5(\alpha_4) \big\},\\
\mathcal{Q}_5^{(\mathbf{C})}&=&\big\{f_0(\alpha_5),f_1(\alpha_5),f_2(\alpha_5),f_3(\alpha_5),f_4(\alpha_5) \big\}.
\end{IEEEeqnarray*}
Accordingly, the answers at servers are computed as
\begin{IEEEeqnarray*}{rCl}
\mathbf{A}_0^{\!\!(\mathbf{C})}\!\!&\!=\!&\!f_1(\!\alpha_0\!)\mathbf{w}_1\!+\!f_2(\!\alpha_0\!)\mathbf{w}_2\!+\!f_3(\!\alpha_0\!)\mathbf{w}_3\!+\!f_4(\!\alpha_0\!)\mathbf{w}_4\!+\!f_5(\!\alpha_0\!)\mathbf{w}_5,\\
\mathbf{A}_1^{\!\!(\mathbf{C})}\!\!&\!=\!&\!f_0(\!\alpha_1\!)\mathbf{w}_0\!+\!f_2(\!\alpha_1\!)\mathbf{w}_2\!+\!f_3(\!\alpha_1\!)\mathbf{w}_3\!+\!f_4(\!\alpha_1\!)\mathbf{w}_4\!+\!f_5(\!\alpha_1\!)\mathbf{w}_5,\\
\mathbf{A}_2^{\!\!(\mathbf{C})}\!\!&\!=\!&\!f_0(\!\alpha_2\!)\mathbf{w}_0\!+\!f_1(\!\alpha_2\!)\mathbf{w}_1\!+\!f_3(\!\alpha_2\!)\mathbf{w}_3\!+\!f_4(\!\alpha_2\!)\mathbf{w}_4\!+\!f_5(\!\alpha_2\!)\mathbf{w}_5,\\
\mathbf{A}_3^{\!\!(\mathbf{C})}\!\!&\!=\!&\!f_0(\!\alpha_3\!)\mathbf{w}_0\!+\!f_1(\!\alpha_3\!)\mathbf{w}_1\!+\!f_2(\!\alpha_3\!)\mathbf{w}_2\!+\!f_4(\!\alpha_3\!)\mathbf{w}_4\!+\!f_5(\!\alpha_3\!)\mathbf{w}_5,\\
\mathbf{A}_4^{\!\!(\mathbf{C})}\!\!&\!=\!&\!f_0(\!\alpha_4\!)\mathbf{w}_0\!+\!f_1(\!\alpha_4\!)\mathbf{w}_1\!+\!f_2(\!\alpha_4\!)\mathbf{w}_2\!+\!f_3(\!\alpha_4\!)\mathbf{w}_3\!+\!f_5(\!\alpha_4\!)\mathbf{w}_5,\\
\mathbf{A}_5^{\!\!(\mathbf{C})}\!\!&\!=\!&\!f_0(\!\alpha_5\!)\mathbf{w}_0\!+\!f_1(\!\alpha_5\!)\mathbf{w}_1\!+\!f_2(\!\alpha_5\!)\mathbf{w}_2\!+\!f_3(\!\alpha_5\!)\mathbf{w}_3\!+\!f_4(\!\alpha_5\!)\mathbf{w}_4.
\end{IEEEeqnarray*}

Remarkably, the answer $\mathbf{A}_n^{(\mathbf{C})}$ from each server $n,n=0,1,2,3,4,5$ is equal to evaluating the following polynomial $h(x)$ of degree $4$ at point $x=\alpha_n$.
\begin{IEEEeqnarray}{l}
h(x)=f_0(x)\mathbf{w}_0+f_1(x)\mathbf{w}_1+f_2(x)\mathbf{w}_2\notag\\
\quad\quad\quad\quad\quad\quad\quad\quad+f_3(x)\mathbf{w}_3+f_4(x)\mathbf{w}_4+f_5(x)\mathbf{w}_5.\notag
\end{IEEEeqnarray}
Therefore, the user can interpolate the polynomial $h(x)$ from any $N-S=5$ server answers.
Finally, evaluating $h(x)$ at $x=\beta_0,\beta_1,\beta_2$ will obtain
\begin{IEEEeqnarray}{rCl}
h(\beta_0)&=&
\left[
\begin{array}{@{}c@{}}
c_{0,0} \\
c_{1,0} \\
\end{array}
\right]\mathbf{w}_0+
\left[
\begin{array}{@{}c@{}}
c_{0,1} \\
c_{1,1} \\
\end{array}
\right]
\mathbf{w}_1+
\left[
\begin{array}{@{}c@{}}
c_{0,2} \\
c_{1,2} \\
\end{array}
\right]
\mathbf{w}_2,\notag\\
h(\beta_1)&=&
\left[
\begin{array}{@{}c@{}}
c_{2,0}\mathbf{w}_0+c_{2,1}\mathbf{w}_1+c_{2,2}\mathbf{w}_2 \\
c_{0,0}\mathbf{w}_3+c_{0,1}\mathbf{w}_4+c_{0,2}\mathbf{w}_5 \\
\end{array}
\right],\notag\\
h(\beta_2)&=&
\left[
\begin{array}{@{}c@{}}
c_{1,0} \\
c_{2,0} \\
\end{array}
\right]\mathbf{w}_3+
\left[
\begin{array}{@{}c@{}}
c_{1,1} \\
c_{2,1} \\
\end{array}
\right]\mathbf{w}_4+
\left[
\begin{array}{@{}c@{}}
c_{1,2} \\
c_{2,2} \\
\end{array}
\right]\mathbf{w}_5.\notag
\end{IEEEeqnarray}

This completes the desired linear computations $\widetilde{\mathbf{C}}\cdot\mathbf{W}$. Moreover, privacy is guaranteed because all the coefficients in \eqref{example:partition} are padded by independently uniformly distributed random noises, and the query received by any $T=1$ server cannot obtain any information about the coefficients \eqref{example:coefficients}.

\subsection{General Description of PMLC Scheme}
Let $K>0,E>0$ and $R\geq 0$ be three integer parameters such that 
\begin{IEEEeqnarray}{c}\label{design:parame}
    K+R\leq N-S-T.
\end{IEEEeqnarray}

Without loss of generality, we assume that each file $\mathbf{w}^{(m)}$ is arranged in a \emph{row} vector of length $L$ for all $m\in[0:M)$.
To reduce the download cost, the file $\mathbf{w}^{(m)}$ is uniformly partitioned into $E$ disjoint pieces $\mathbf{w}^{(m)}_{0},\ldots,\mathbf{w}^{(m)}_{E-1}$, each with a length of ${L}/{E}$, i.e.,
\begin{IEEEeqnarray}{c}\label{vector:partition}
    \mathbf{w}^{(m)}=\left[ \mathbf{w}^{(m)}_{0},\ldots,\mathbf{w}^{(m)}_{E-1} \right], \quad\forall\,m\in[0:M).
\end{IEEEeqnarray}
Then the $M$ files $\mathbf{w}^{(0)},\ldots,\mathbf{w}^{(M-1)}$ stored at each server can be equivalently written into a matrix $\mathbf{W}$ of dimensions $ME\times{L}/{E}$, given by
\begin{IEEEeqnarray}{rCl}\label{file:symbols}
\mathbf{W}=\left[
  \begin{array}{c}
\mathbf{w}_{0} \\
\vdots \\
\mathbf{w}_{M-1} \\
\vdots\\
\mathbf{w}_{M(E-1)} \\
\vdots \\
\mathbf{w}_{ME-1} \\
\end{array}
\right]=\left[
  \begin{array}{c}
\mathbf{w}^{(0)}_{0} \\
\vdots \\
\mathbf{w}^{(M-1)}_{0} \\
\vdots\\
\mathbf{w}^{(0)}_{E-1} \\
\vdots \\
\mathbf{w}^{(M-1)}_{E-1} \\
\end{array}
\right],
\end{IEEEeqnarray}
where $\mathbf{w}_{eM+m}=\mathbf{w}^{(m)}_{e}$ for any $m\in[0:M)$ and $e\in[0:E)$.

Let $\widetilde{\mathbf{C}}$ be an expanding matrix of the interested coefficients $\mathbf{C}$ defined in \eqref{coefficient:matrix}, given by
\begin{IEEEeqnarray}{c}\notag
\widetilde{\mathbf{C}}=\mathrm{diag}\Big(\underbrace{\mathbf{C},\mathbf{C},\ldots,\mathbf{C}}_{E} \Big),
\end{IEEEeqnarray}
whose dimensions are $PE\times ME$.
Accordingly, the $P$ desired linear computations $\widetilde{\textbf{w}}^{(0)},\ldots,\widetilde{\textbf{w}}^{(P-1)}$ in \eqref{desired:computation} can be completed by obtaining the computational results
\begin{IEEEeqnarray}{c}\label{Desired:computation}
    \widetilde{\mathbf{C}}\cdot\mathbf{W}.
\end{IEEEeqnarray}

To reduce upload cost and server computation complexity, we horizontally divide the matrix $\widetilde{\mathbf{C}}$ into $K$ disjoint sub-matrices $\widetilde{\mathbf{C}}^{0},\!\ldots\!,\widetilde{\mathbf{C}}^{K\!-\!1}$, each of dimensions $\frac{PE}{K}\!\times\! ME$, i.e.,
\begin{IEEEeqnarray}{rCl}\label{expanding:matrix}
\widetilde{\mathbf{C}}=\left[
  \begin{array}{c}
\widetilde{\mathbf{C}}^0 \\
\vdots \\
\widetilde{\mathbf{C}}^{K-1}
\end{array}
\right].
\end{IEEEeqnarray}


Assume the size of the finite field $\mathbb{F}_q$ satisfies $q\!\geq\!N\!+\!K\!+\!T$. Let $\{\beta_{k},\alpha_n\!:\!k\!\in\![0\!:\!K+T),n\!\in\![0\!:\!N)\}$ be $N\!+\!K\!+\!T$ pairwise distinct elements from $\mathbb{F}_q$.
Denote by $\widetilde{\mathbf{c}}^{k}_{\ell}$ the $\ell$-th \emph{column} of the sub-matrix $\widetilde{\mathbf{C}}^k$ for any given $k\in[0:K)$ and $\ell\in[0:ME)$.

To provide a privacy guarantee of $\mathbf{C}$ against any $T$ colluding servers, we employ $T$ random noises $\mathbf{z}_{\ell}^{K},\ldots,\mathbf{z}_{\ell}^{K+T-1}$ to mask the $K$ coefficient vectors $\widetilde{\mathbf{c}}^{0}_{\ell},\ldots,\widetilde{\mathbf{c}}^{K-1}_{\ell}$ in the $\ell$-th column of matrix $\widetilde{\mathbf{C}}$ for all $\ell\in[0:ME)$, where each of random noises is a column vector of length ${PE}/K$ and its elements are all distributed independently and uniformly over $\mathbb{F}_q$. This is achieved by using Lagrange polynomials to interpolate these noises and coefficient vectors. Particularly, to further reduce the upload cost and server computation cost, we additionally force the evaluations of each interpolating polynomial to be $\mathbf{0}_{PE/K}$ at $R$ specific points, where $\mathbf{0}_{PE/K}$ is a column vector of length $PE/K$ with all elements being $0$.
Specifically, choose a polynomial $f_{\ell}(x)$ of degree at most $K+T+R-1$ for every $\ell\in[0:ME)$ such that
\begin{IEEEeqnarray}{rcl}
f_{\ell}(\beta_{k})&=&\left\{
\begin{array}{@{}ll}
\widetilde{\mathbf{c}}^{k}_{\ell} ,&\forall\, k\in[0:K)\\
\mathbf{z}_{\ell}^{k},&\forall\, k\in[K:K+T)
\end{array}\right., \label{interpolate} \\
f_{\ell}(\alpha_{(\ell-r)_{N}})&=&\mathbf{0}_{\frac{PE}{K}}, \quad\forall\,r\in[0:R). \label{force:zero}
\end{IEEEeqnarray}  

The evaluations of the $ME$ polynomials $\{f_{\ell}(x):\ell\in[0:ME)\}$ at point $x=\alpha_n$ are given by
\begin{IEEEeqnarray}{c}\label{evaluation}
    \left\{f_{\ell}(\alpha_n):\ell\in[0:ME) \right\},\quad\forall\,n\in[0:N).
\end{IEEEeqnarray}

It is not difficult to demonstrate that 
$\{\ell:(\ell-r)_{N}=n,r\in[0:R),\ell\in[0:ME)\}=\{(n+r)_N+sN:r\in[0:R),s\in[0:{ME}/{N})\}$ for any given $n\in[0:N)$.
Thus by \eqref{force:zero},
\begin{IEEEeqnarray}{c}\label{zeros:valuations}
f_{(n+r)_N+sN}(\alpha_n)\!=\!\mathbf{0}_{\frac{PE}{K}},\;\;\;\forall\, r\!\in\![0\!:\!R),s\!\in\![0\!:\!\frac{ME}{N}). \IEEEeqnarraynumspace
\end{IEEEeqnarray}
Then the query $\mathcal{Q}_{n}^{(\mathbf{C})}$ sent to server $n\in[0:N)$ is comprised of the nonzero evaluations in \eqref{evaluation}, i.e.,
\begin{IEEEeqnarray}{c}\label{PMLC:query}
   \mathcal{Q}_{n}^{(\mathbf{C})}\!=\!\left\{\! 
   f_{(n+r)_N+sN}(\alpha_n):r\!\in\![R\!:\!N),s\!\in\![0\!:\!\frac{ME}{N})\!\right\}. \IEEEeqnarraynumspace
\end{IEEEeqnarray}

Upon receiving the query $\mathcal{Q}_{n}^{(\mathbf{C})}$, the server $n$ generates an answer by computing the products of the nonzero evaluations in \eqref{PMLC:query} and the corresponding data vectors in \eqref{file:symbols} and then summing them, given by\footnote{
Note that in practice, the PMLC scheme is public to the servers, except for the random noises used. Thus upon receiving the query, each server $n\in[0:N)$ only needs to additionally know the hyperparameters $R$ and $E$ to be able to generate the answer in \eqref{answer:data}. 
Moreover, the communication cost for sending $R$ and $E$ is negligible.}
\begin{IEEEeqnarray}{c}\label{answer:data}
\mathbf{A}_{n}^{(\mathbf{C})}\!=\!\sum\limits_{r\in[R:N)}\sum\limits_{s\in[0:\frac{ME}{N})}\!\!
   f_{(n+r)_N+sN}(\alpha_n)\!\cdot\! \mathbf{w}_{(n+r)_N+sN}. \IEEEeqnarraynumspace
\end{IEEEeqnarray}

Next, we describe the decoding steps.
From \eqref{zeros:valuations}, the answer $\mathbf{A}_{n}^{(\mathbf{C})}$ can be equivalently expressed as
\begin{IEEEeqnarray}{rCl}\label{answer:poly}
\mathbf{A}_{n}^{(\mathbf{C})}&=&\sum\limits_{r\in[0:N)}\sum\limits_{s\in[0:\frac{ME}{N}\!)}\!
   f_{(n+r)_N+sN}(\alpha_n)\cdot\mathbf{w}_{(n+r)_N+sN} \notag \\
   &=&\sum\limits_{\ell\in[0:ME)}
   f_{\ell}(\alpha_n)\cdot \mathbf{w}_{\ell}.
\end{IEEEeqnarray}

Denote the answer polynomial $h(x)$ by
\begin{IEEEeqnarray}{c}
   h(x)=\sum\limits_{\ell\in[0:ME)}
   f_{\ell}(x)\cdot \mathbf{w}_{\ell}.
\end{IEEEeqnarray}
Apparently, the answer $\mathbf{A}_{n}^{(\mathbf{C})}$ is an evaluation of the polynomial $h(x)$ at point $x=\alpha_n$ for any $n\in[0:N)$. Moreover, by \eqref{interpolate} and \eqref{force:zero}, $h(x)$ is a polynomial of degree at most $K+T+R-1$. Since the elements $\{\alpha_n:n\in[0:N)\}$ are all distinct on $\mathbb{F}_q$, the answer polynomial $h(x)$ can be recovered from any $N-S$ server answers by \eqref{design:parame} and Lagrange interpolating theorem.

Furthermore, evaluating $h(x)$ at point $x=\beta_{k}$ for any $k\in[0:K)$ will obtain
\begin{IEEEeqnarray}{c}\label{answer:evalu}
   h(\beta_k)=\sum\limits_{\ell\in[0:ME)}
   \widetilde{\mathbf{c}}^{k}_{\ell}\cdot \mathbf{w}_{\ell}=\widetilde{\mathbf{C}}^k\cdot\mathbf{W},
\end{IEEEeqnarray}
where the first equation is due to \eqref{interpolate}, and the last equation is due to \eqref{file:symbols} and \eqref{expanding:matrix}.
By jointing $\{\widetilde{\mathbf{C}}^k\cdot\mathbf{W}:k\in[0:K)\}$, we obtain the computational results $\widetilde{\mathbf{C}}\cdot\mathbf{W}$ in \eqref{Desired:computation}.

As a result, the user can decode the desired linear computations from any $N-S$ answers.

\subsection{Performance Analysis}
In this subsection, we analyze the communication costs, computational complexities, and privacy guarantee of the proposed PMLC scheme.

\begin{Lemma}[\!\!\cite{von2013modern}]\label{Lemma:Evaluation}
The evaluation or interpolation of an $n$-th degree polynomial at $n+1$ arbitrary points can be done in ${{O}}(n(\log n)^2\log\log n)$ arithmetic operations. 
\end{Lemma}
\subsubsection*{Upload Cost}
From \eqref{interpolate} and \eqref{force:zero}, the evaluation of the polynomial $f_{\ell}(x)$ at any given point is a \emph{column} vector of length $\frac{PE}{K}$ for any $\ell\in[0:ME)$. The query sent to each server consists of $\frac{(N-R)ME}{N}$ nonzero evaluations by \eqref{PMLC:query}. Thus the upload cost for $N$ queries is $\frac{(N-R)E^2MP}{K}$.

\subsubsection*{Download Cost} By \eqref{vector:partition} and \eqref{file:symbols}, $\mathbf{w}_{\ell}$ is a \emph{row} vector of length $\frac{L}{E}$ for any $\ell\in[0:ME)$. Thus the answer \eqref{answer:data} at each server is a matrix of dimensions $\frac{PE}{K}\times\frac{L}{E}$. Accordingly, the download cost from any $N-S$ responsive servers is $\frac{(N-S)PL}{K}$.

\subsubsection*{Query Complexity}
From \eqref{PMLC:query}, the queries sent to servers are constructed by evaluating $ME$ polynomials of each degree $K+T-R-1<N-R$ at $N-R$ points for $\frac{PE}{K}$ times. This achieves a complexity of $O(\frac{E^2MP(N-R)(\log(N-R))^2\log\log(N-R)}{K})$ by Lemma \ref{Lemma:Evaluation}.

\subsubsection*{Server Computation Complexity}
By \eqref{answer:data}, the answer at each server can be viewed as computing the product of a column vector of length $\frac{PE}{K}$ and a row vector of length $\frac{L}{E}$ for $\frac{(N-R)ME}{N}$ times and then summing these computation results, which incurs a complexity of $O(\frac{E(N-R)MPL}{NK})$.

\subsubsection*{Decoding Complexity}
In the decoding phase \eqref{answer:poly}--\eqref{answer:evalu}, the user first interpolates an answer polynomial with degree $K+T+R-1<N-S$ and dimensions $\frac{PE}{K}\times\frac{L}{E}$, and then evaluates this polynomial at $K<N-S$ points. This can be completed within the complexity of $O(\frac{PL(N-S)(\log (N-S))^2\log\log (N-S)}{K})$ by Lemma \ref{Lemma:Evaluation}.

\subsubsection*{Privacy}
Note from \eqref{interpolate} and \eqref{force:zero} that the $T$ evaluations $f_\ell(\alpha_{n_0}),\ldots,f_\ell(\alpha_{n_{T-1}})$  are padded with $T$ independently uniformly random noises for all $\ell\in[0:KE)$ and any given $\mathcal{T}=\{n_0,\ldots,n_{T-1}\}\subseteq[0:N)$ with $|\mathcal{T}|\leq T$. In essence, this matches the encryption technique of one-time pad (OTP) \cite{katz2007introduction}. Thus the queries \eqref{PMLC:query} sent to the servers $\mathcal{T}$ provide no information about the coefficients $\mathbf{C}$, achieving the privacy constraint in \eqref{Infor:priva cons}.


\section*{Acknowledgment}
This work was supported in part by the National Natural Science Foundation of China (NSFC) under Grant 62011530134 and 62371401, and was supported in part by the Fundamental Research Funds for the Central Universities under Grant 2682024CX030.

\bibliographystyle{ieeetr}
\bibliography{reference.bib}

\end{document}